\documentclass[prl,twocolumn,aps,showpacs,superscriptaddress,amssymb]{revtex4}
\usepackage{graphicx}
\usepackage{epsfig}
\usepackage[usenames,dvipsnames]{xcolor}
\usepackage{ulem}
\usepackage{natbib}
\usepackage{hyperref}
\usepackage{color}

\usepackage{color}

\def\lsim{\lesssim}                
\def\gsim{\gtrsim}                

\newcommand{\nn}[1]{{\langle{#1}\rangle}}

\newcommand{\Eqref}[1]{Eq. (\ref{#1})}

\newcommand{\Fref}[1]{Fig.~\ref{#1}}

\newcommand{\mc}{\mathcal}
\newcommand{\be}{\begin{eqnarray}}
\newcommand{\ee}{\end{eqnarray}}
\renewcommand{\v}[1]{\mathbf{#1}}

\begin{document}

\title{Adiabatic loading of one-dimensional SU($N$) alkaline earth fermions in optical lattices}

\author{Lars Bonnes}
\affiliation{Institute for Theoretical Physics, University of Innsbruck, A-6020 Innsbruck, Austria.}
\email{lars.bonnes@uibk.ac.at}

\author{Kaden R. A. Hazzard}
\affiliation{JILA, NIST and University of Colorado, and Department of Physics, University of Colorado, Boulder, Colorado 80309-0440, USA.}

\author{Salvatore R. Manmana}
\affiliation{JILA, NIST and University of Colorado, and Department of Physics, University of Colorado, Boulder, Colorado 80309-0440, USA.}

\author{Ana Maria Rey}
\affiliation{JILA, NIST and University of Colorado, and Department of Physics, University of Colorado, Boulder, Colorado 80309-0440, USA.}

\author{Stefan Wessel}
\affiliation{Institute for Theoretical Solid State Physics, JARA-FIT,  and JARA-HPC, RWTH Aachen University, Otto-Blumenthal-Str. 26, D-52056 Aachen, Germany.}

\date{\today}

\begin{abstract}
Ultracold fermionic alkaline earth atoms confined in optical lattices realize Hubbard models with internal $\mathrm{SU}(N)$ symmetries, where $N$ can be as large as ten.
Such systems are expected to harbor exotic magnetic physics at temperatures below the  superexchange energy scale.
Employing quantum Monte Carlo simulations to access the low-temperature regime of one-dimensional chains, we show that after adiabatically loading a weakly interacting gas into the strongly interacting regime of an optical lattice,
the final temperature decreases with increasing $N$.
Furthermore, we estimate the temperature scale required to probe correlations associated with low-temperature SU($N$) magnetism.
Our findings  are encouraging for the exploration of exotic large-$N$ magnetic states in ongoing experiments.
\end{abstract}


\maketitle


Ultracold fermionic alkaline earth atoms confined in optical lattices realize an important, tunable generalization of the Hubbard model, widely used to model strongly correlated electrons~\cite{gorshkov10,wu:exact_2003,cazalilla_ultracold_2009,
fukuhara_mott_2009,sugawa:interaction_2011}.    Within this generalization, the conventional spin-1/2 SU(2) symmetry is enhanced to an SU($N$) symmetry. $N=2I+1$ is  determined  by the nuclear spin, $I$,  which  varies in alkaline earths from $I=1/2$ to $I=9/2$ depending on the atomic species.  The SU($N$) symmetry arises because the  electronic degrees of freedom have neither spin nor
orbital angular momentum  (due to the closed-shell structure) and  thus decouple from the nuclear spin. In the electronic ground state the symmetry has been  theoretically predicted to hold to an accuracy of $10^{-9}$~\cite{gorshkov10} and
experiments have constrained deviations to be less than $5\times 10^{-4}$~\cite{stellmer11}.  Despite the large $I$, quantum fluctuations remain important due to the enhanced symmetry, giving rise to  magnetic frustration and
 exotic ground states, such as valence-bond solids~\cite{read_valence-bond_1989,nonne11,bauer12,affleck_large-n_1988,honerkamp_ultracold_2004}, exotic spin orderings~\cite{toth_three_2010} or (chiral) spin liquids~\cite{hermele09,hermele_topological_2012}
 in addition to the plethora of potential phases for the conventional $N=2$ Hubbard model, such as antiferromagnets, d-wave superconductors, and nematic states. The possibility of mimicking such exciting many-body physics, as well as potential applications to atomic clocks~\cite{ido_recoil-free_2003,ludlow_sr_2008,lemke:spin_2009,
derevianko_physics_2011}, measurements of fundamental constants~\cite{kotochigova_prospects_2009}, and quantum information processing~\cite{daley_quantum_2011}, has stimulated substantial experimental progress~\cite{takasu_spin-singlet_2003,fukuhara_bose-einstein_2007,
fukuhara_mott_2009,kraft_bose-einstein_2009,stellmer_bose-einstein_2009,
de_escobar_bose-einstein_2009,mickelson_bose-einstein_2010,
desalvo_degenerate_2010,tey_double-degenerate_2010,
taie_realization_2010,stellmer:creation_2012}.

Although cold atom experiments routinely reach nanokelvin temperatures, it is an ongoing effort to achieve temperatures and entropies sufficiently low to see superexchange driven magnetic many-body physics~\cite{mckay_cooling_2011,joerdens08,schneider08a,sugawa:interaction_2011}.
Hazzard \textit{et al.} showed via a high temperature series expansion (HTSE) that for final temperatures $T\gsim t$, with $t$ the tunneling rate in the lattice, the temperatures reached after adiabatic loading from experimentally realistic initial conditions decreases with increasing $N$~\cite{hazzard10}.
Closely related is the finding that the entropy of Mott states increases with $N$ even faster than the initial weakly interacting gas entropy.
The question of the behavior below $t$, and especially below the magnetic exchange energy scale $J \sim t^2/U$ where $U$ is the on-site interaction, has remained open even though this is one relevant for exploring exotic
$\mathrm{SU}(N)$ magnetism.
Two issues are particularly relevant in this regime: how does $N$ affect (i) the temperature reached by adiabatic loading and (ii) the physical properties, such as correlation functions?

Here, we address both questions in one dimensional systems using quantum Monte Carlo calculations.
We show that also for $T<t^2/U$ the temperatures reached by adiabatic loading decrease with increasing $N$.
This decrease occurs even relative to the temperature scales of interesting physics, for example the onset of Luttinger liquid behavior, magnetic correlations, or ground state-like correlations.

The Hamiltonian of the SU($N$) Hubbard model describing alkaline earth atoms in optical lattices is~\cite{gorshkov10}
\begin{equation}
\mathcal{H}=-t\sum_{i,\alpha} \left( f_{\alpha,i}^\dagger f_{\alpha,i+1} + \mathrm{H.c.} \right)
+ \frac{U}{2} \sum_{i,\alpha \ne \beta} n_{i}^\alpha n_{i}^\beta,
\label{eq:ham}
\end{equation}
where $f_{\alpha,i}^\dagger$ ($f_{\alpha,i}$) are creation (annihilation) operators for fermions of flavor $\alpha$ at site $i$, and $n_{i}^\alpha=f_{\alpha,i}^\dagger f_{\alpha,i}$. At filling $1/N$ (density $n=1$)
local moment (Heisenberg) quantum magnetism arises in the strong interaction limit,
where second order processes lead to a flavor exchange interaction scale $J\sim t^2/U$.
We employ quantum Monte-Carlo (QMC) to obtain the low temperature thermodynamic properties of one-dimensional lattice alkaline earth atoms described by \Eqref{eq:ham} in the relevant regime $T \ll t \ll U$,  much lower than the temperature scales accessible within the HTSE.

\begin{figure}
 \begin{center}
  \includegraphics[width=\columnwidth]{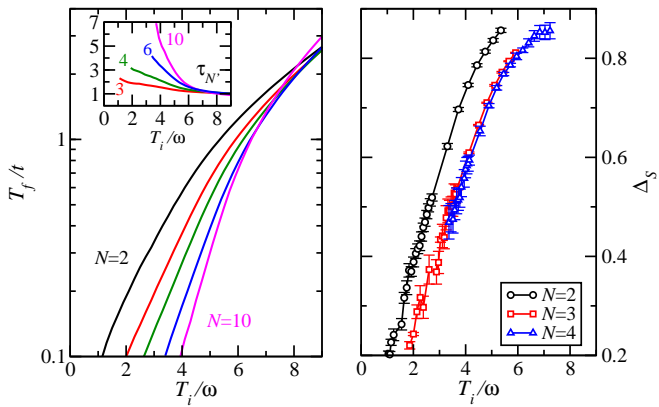}
 \end{center}
\caption{(Color online)
{
{\it Left:} Final vs. initial temperature for the adiabatic loading scheme for $N=2$, 3, 4, 6 and 10 (from left to right) inside a harmonic trap with trap frequency $\omega=2\pi \times 90$Hz and for $U/t=8$. 
One can alternatively translate the initial temperature into units of Fermi temperature $T_F$, using
$T_i/T_F=\{2.8,3.2,3.5,4.1,4.8\}\times10^{-2}(T_i/\omega)$ for $N=\{2,3,4,6,10\}$.
The inset shows the temperature decrease relative to the $N=2$ Hubbard model, $\tau_{N'}=T_f(N=2)/T_f(N=N')$, for (from bottom to top) $N'=3$ (red), $N'=4$ (green), $N'=6$ (blue) and $N'=10$ (magenta) vs. $T_i/\omega$.
\it Right:} Distance measure $\Delta_S$ of the magnetic correlations relative to the ground state values for a homogeneous system with $U/t=8$ at $n=1$.  
}
\label{fig:PlotLoading}
\end{figure}

These thermodynamic properties allow us to calculate the temperatures that experiments can achieve using the standard experimental adiabatic lattice loading protocol: the systems are prepared by starting with a weakly interacting gas and slowly turning on the lattice~\cite{joerdens08,schneider08a}.  Ideally, this is done sufficiently slowly to maintain adiabaticity, minimizing the increase in entropy.  The final temperature in the adiabatic limit in this sense provides a lower bound to the experimentally achievable temperatures, and in practice the adiabatic bound is frequently an accurate approximation~\cite{joerdens08,schneider08a,natu:local_2011}.

In the following, we  consider loading a three dimensional  ($d=3$) trapped gas (no lattice) into a two dimensional array of independent one dimensional lattices.
Fig.~\ref{fig:PlotLoading} (left panel) shows our central result, the final temperature $T_f$ of the harmonically trapped lattice system with  fixed trapping frequency $\omega$ as a function of the initial temperature $T_i$ before applying the lattice~\cite{Note1}.  
We use realistic experimental particle number and trap frequencies, similar to Ref.~\cite{hazzard10}: $\mc N=1.5\times 10^4$ and $\omega=2\pi \times 90$Hz.
For loading ${}^{173}$Yb in a $266\,\mathrm{nm}$ lattice, our choice of $U$ corresponds to the lattice depth required to obtain $U/t=8$.
We observe from Fig.~\ref{fig:PlotLoading} that the final temperature achieved for $T_i/\omega < 9$ significantly decreases with increasing $N$. This includes the lowest temperatures, an order of magnitude lower than where the HTSE is applicable, and is encouraging for experimentally achieving SU($N$) quantum magnetism.
For instance,  at the relatively warm temperature $T_i/\omega=4$ (some experiments are already at even lower temperatures) the final temperature is already decreased by factors of 1.7 ($N=3$), 1.85 ($N=4$), 3.25 ($N=6$) and 5.6 ($N=10$) compared to the conventional $N=2$ case. We note that even this is a pessimistic estimate since these ratios compare different $N$ fixing $T_i/\omega$. Experimentally, $T_i$ will likely decrease with $N$, as discussed in Ref.~\citealp{hazzard10}.
The reason why  $T_f$ decreases with $N$, for fixed initial $T_i$, is the scaling of the initial and final states' entropy with $N$.
At very low temperature $T\ll t^2/U$, the Mott insulator possesses $N-1$ gapless channels, and the metal possesses $N$~\cite{lee_low-temperature_1994,manmana11a}. The increasing number of gapless excitations implies that the entropy growth at fixed temperature is faster than the initial state's $N^{1/3}$.  Thus, the final temperatures will decrease with increasing $N$.  This is even more favorable than at high temperatures $T\gsim t$, where the Mott insulator accommodates an entropy $S \sim\log N$~\cite{hazzard10}, as can be seen in Fig.~\ref{fig:PlotLoading}.
However, even for $T>t$, for $N\lesssim 20$ the logarithmic term grows \textit{fast} enough to compensate for the $N^{1/3}$ in the initial state, leading to colder final states with increasing $N$.

The enhanced cooling effect has in deed been observed (at elevated temperatures) for SU(6) fermions in recent experiments on a ${}^{173}$Yb gas in a three-dimensional optical lattice~\cite{taie_2012}.
In particular, their findings are in good agreement with the HTSE results (valid for their experiments), confirming the validity of the adiabaticity assumption for their experiments.

We now describe the procedure used to obtain Fig.~\ref{fig:PlotLoading}.
Assuming adiabaticity, the initial temperature $T_i$ and particle number ${\mc N}_i$ in the absence of the lattice uniquely determine the final temperature $T_f$ through the conservation of entropy and particle number. $T_f$ can be calculated in terms of the experimentally measurable $T_i$ and ${\mc N}_i$ once one knows how the particle number and entropy of the  system in a lattice depend on temperature $T$ and the chemical potential $\mu$, ${\mc N}_f(T,\mu)$ and $S_f(T,\mu)$, as follows:
We calculate $T_f$ and  $\mu_f$ by solving particle number and entropy conservation equations, ${\mc N}_f(T_f,\mu_f)={\mc N}_i(T_i,\mu_i)$ and $S_f(T_f,\mu_f)=S_i(T_i,\mu_i)$, for $T_f$ and $\mu_f$ in terms of the initial temperature $T_i$ and chemical potential $\mu_i$ of the weakly interacting gas. In practice, since experimentalists measure the initial particle number ${\mc N}_i$ rather than $\mu_i$, we solve for $\mu_i$ in terms of $T_i$ and ${\mc N}_i$ using ${\mc N}_i(\mu_i,T_i)={\mc N}_i$.  The conservation equations become
\be
{\mc N}_f(T_f,\mu_f) = {\mc N}_i,
\:\:
S_f(T_f,\mu_f) =S_i(T_i, \mu_i({\mc N}_i,T_i)). \label{eq:adiab-load}
\ee
The functions ${\mc N}_i(T,\mu)$ and $S_i(T,\mu)$ are those of a gas in a harmonic trapping potential $V(r)=m\omega^2r^2/2$.  For relevant experimental initial conditions, the temperature is low compared to the Fermi temperature, interactions are weak, and the number of particles is large, so the semiclassical approximation to the non-interacting degenerate Fermi gas is accurate, giving~\cite{hazzard10}
$
{\mc N}_i(\mu,T) = (N/d!)(\mu/\omega)^d
$
and
$
S_i(\mu,T)  = \frac{T_i}{\omega}\frac{N \pi^2}{3(d-1)!}\left(\frac{\mu}{\omega}\right)^{d-1}.
$
We consider dimension $d=3$.

\begin{figure}
 \begin{center}
  \includegraphics[width=\columnwidth]{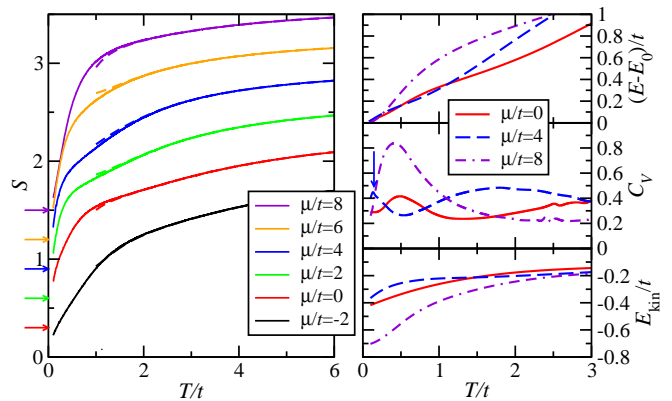}
 \end{center}
\caption{(Color online)
\textit{Left:} Entropy $S$ per site for $N=3$, $U/t=8$ at fixed chemical potentials from $\mu/t=-2$ (bottom) to $\mu/t=8$ (top) obtained from QMC for a homogeneous system. We note, that the system forms a Mott insulating ground state with density $n=1$ for $\mu/t\approx 1.5$ to $5$.
Dashed lines down to $T/t=1$ denote HTSE results from Ref.~\citealp{hazzard10}.
For clarity, an offset was added to the entropies, indicated by  the arrows.
\textit{Right top:} Temperature dependence of the energy per site $E$ relative to the extrapolated ground state energy $E_0$, from QMC.
\textit{Right middle:} Temperature dependence of the specific heat $C_V$~\cite{Note4}. The arrow indicates the low-$T$ peak in $C_V$ in the Mott regime.
\textit{Right bottom:} Temperature dependence of the kinetic energy $E_{kin}$, from QMC.
Statistical errors are below the line width.
}
\label{fig:Entropies}
\end{figure}

We obtain the functions ${\mc N}_f(T,\mu)$ and $S_f(T,\mu)$
in two steps.  First, to compute the total particle and entropy in the trap we apply the local density approximation (LDA)~\cite{pethick02}: we obtain the properties at position $\v{r}$ from those of a homogeneous system at a chemical potential $\mu(\v{r})=\mu-V(\v{r})$.
Due to the large particle number (about 30 particles in the central tube at $T=0$), 
this approximation will be accurate.
Second, to obtain the homogeneous system properties used in the LDA, we use sign-problem-free QMC simulations~\cite{Note2} within the stochastic series expansion (SSE) framework~\cite{sandvik99b,syljuasen02,alet05}.  We first calculate the density and entropy for finite systems up to $L=100$ sites, with open boundaries within the grand-canonical ensemble for various values of the chemical potential $\mu$. We find finite size effects to be negligible on these lattice sizes for the quantities of interest below.
The entropy is obtained by a standard thermodynamic integration of the energy $E$.
In fact, the HTSE agrees well with the QMC data down to $T/t \approx 2$, as illustrated in \Fref{fig:Entropies}, and we thus used the HTSE results at
$T/t=10$ as the high-temperature end point for the thermodynamic integration.
Increasingly dense temperature grids are required to perform
this integration down to lower temperatures, and we were able to
perform this procedure down to $T/t=0.1$. We find that our data indeed
connects to the  $T\rightarrow 0$ limit, where $S$ scales linear in
$T$~\cite{lee_low-temperature_1994}.
We finally compared our QMC results to ground state properties~\cite{manmana11a},
obtained using the density matrix renormalization group (DMRG)~\cite{white92,white93,peschel99,schollwoeck05}. In particular,
the extrapolated ground state energies agree with DMRG results within the statistical error bars.
Finally, we calculate tables of ${\mc N}_f$ and $S_f$ for a dense grid of $\mu$ and $T$, construct interpolating functions of this data supplemented, for very negative $\mu$, with the virial expansion that accurately describes the gas in this regime. 
The adiabatic loading equations Eqs.~(\ref{eq:adiab-load}) are then solved numerically to obtain the results in Fig.~\ref{fig:PlotLoading}.

Having determined achievable temperatures, we next turn to correlation functions that are indicative for low-energy SU($N$) magnetism in these systems.
The relevant spin-spin correlation function is related to the equal and unequal flavor density-density correlation functions~\cite{manmana11a}
\begin{equation}
\label{eq:spincorr}
\mathcal{S}(|i-j|)=\frac{1}{N}\sum_\alpha\nn{n_i^\alpha n_j^\alpha} - \frac{1}{N(N-1)}\sum_{\alpha\neq \beta} \nn{n_i^\alpha n_j^\beta}.
\label{eq:corrSpin}
\end{equation}
This can be measured via, for example, Bragg spectroscopy~\cite{carusotto:bragg_2006}.
To compare finite-temperature and ground state correlations (obtained from DMRG), we define an appropriate
distance measure that accounts also for the overall magnitude of the ground state correlations~\cite{manmana11a},
\begin{equation}
\Delta_\mathcal{S}(T)= \frac{\sqrt{\sum_{r=1}^{L} (\mathcal{S}(r,T) - \mathcal{S}(r)_\mathrm{DMRG})^2}}
{\sqrt{\sum_{r=1}^L \mathcal{S}(r)^2_\mathrm{DMRG}}}. \label{eq:Delta-size}
\end{equation}
This quantity is sensitive to both long range and short range magnetic correlations~\cite{greif11,gorelik12}.
The right panel of 
Fig.~\ref{fig:PlotLoading} shows $\Delta_\mathcal{S}$ for $n=1$ and $U/t=8$ as a function of $T_i/\omega$, using $T_f/t$ determined from the lattice loading~\cite{Note3}. This shows that increasing $N$ brings the system not only to lower values of $T$, but also into a region where ground-state-like correlations are more developed.

\begin{figure}
 \begin{center}
  \includegraphics[width=\columnwidth]{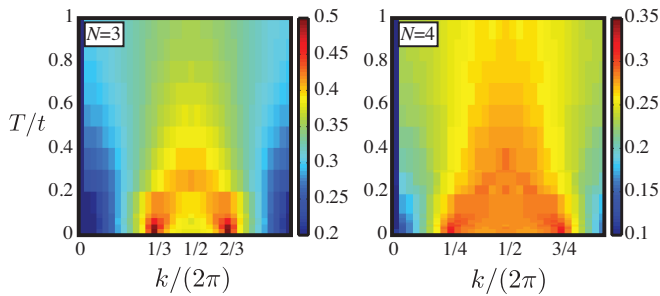}
 \end{center}
\caption{(Color online)
Spin structure factor $\tilde \mathcal{S}(k)$ for a homogeneous system with density $n=1$ for $N=3$ (left) and 4 (right), $U/t=8$ as a function of the final temperature $T/t$ (here, $L=48$).
The characteristic $2k_F$ peaks below $T/t \approx 0.15$ signal the onset of magnetic correlations.
}
\label{fig:SpinSpin}
\end{figure}

A hallmark of quantum magnetism in $N$-component Luttinger liquids (LL) are $2k_F=2\pi  n/ N$ oscillations in the spin correlation functions,  with $n$ the density~\cite{assaraf99,manmana11a}.
Figure \Fref{fig:SpinSpin} shows the spin structure factor $\tilde \mathcal{S}(k)$, the Fourier transformation of $\mathcal{S}(r)$,
for $N=3$ and 4 at $U/t=8$  and fixed filling of $1/N$ ($n=1$).
At low temperatures $T/t \lesssim 0.1$, the $2 k_F$ peak appears, as expected from LL theory.
This is associated with a maximum in the specific heat $C_V$
\cite{Note4}
shown in \Fref{fig:Entropies}, marking the onset of SU($N$) Heisenberg physics~\cite{kawakami89}.
At  high temperatures $T/t \gsim 1$, $\tilde \mathcal{S}(k)$ instead becomes featureless.
However, in an intermediate regime $0.5 \lesssim T/t \lesssim 1$, a broad peak at $k=\pi$ is observed, which shifts towards $2k_F$ as the temperature is lowered  towards $T \approx t^2/U$: at these temperatures, nearest neighbor correlations emerge.
While this regime is not representative for the ground state behavior, the short-ranged magnetic correlations determine the main features of the structure factors, and experiments reaching this temperature regime should be able to investigate magnetic behavior.
In \Fref{fig:Entropies}, the entropy and kinetic energy similarly display the onset of the magnetic exchange region as a kink and large decrease, respectively.
This happens around $T/t\approx 0.5$.
For $N=2$, we have checked that our data are consistent with former studies of the $\mathrm{SU}(2)$ Hubbard model~\cite{kawakami89,sandvik92}.

For intermediate temperature $t^2/U \ll T \ll t$, one can ask whether the system realizes a SU($N$) generalization of the spin-incoherent Luttinger liquid (siLL)~\cite{fiete_colloquium_2007,feiguin_spectral_2010}
where the charge degress of freedom show LL-like behaviour with a simultaneous absence of significant spin correlations.
Here, we find no clear signatures of siLL in the density structure factor near $k=\pi n$ (not shown).  We leave a more in-depth study for future work, where it will likely be beneficial to examine the momentum distribution $n(k)$, as was done for the SU(2) case~\cite{feiguin_spectral_2010}.

\textit{Summary.---} We studied the one-dimensional SU($N$) Fermi-Hubbard model at finite temperatures employing numerically exact quantum Monte Carlo simulations.  We calculated the density and entropy as functions of chemical potential and temperature and used this to determine final temperatures of the lattice system after adiabatically loading an optical lattice from a degenerate gas. We found substantial decreases of the final temperature with increasing $N$, even down to low temperatures $T\lsim t^2/U\ll t$. Together with our results for the temperature dependence of correlation functions, we envisage that it should be possible to explore features of SU($N$) quantum magnetism in ongoing experiments with alkaline earth atoms.

\section*{Acknowledgements}
We acknowledge useful discussions with G. Chen and M. Hermele.
This work was supported by the Austrian Ministry of Science BMWF as part of the UniInfrastrukturprogramm of the Forschungsplattform Scientific Computing at LFU Innsbruck, by the NSF-PIF and NSF-PFC grants, by the AFOSR and by the  ARO with funding from the DARPA-OLE program.
We also acknowledge allocation of CPU time from NIC J\"ulich where parts of the calculations were performed. KH thanks the NRC for support.
KH also thanks the Aspen Center for Physics, which is supported by the NSF, for its hospitality while a portion of this work was carried out.

\textit{Note added.---} Recently, Messio and Mila explored the $N$-dependence of the entropy and correlations in the Heisenberg limit of the SU($N$) systems considered here~\cite{messio:entropy_2012}; their results are in accord with our finding for the Hubbard model, obtained at $U/t=8$. 

\sloppy
\bibliographystyle{prsty}

\end{document}